\documentstyle[eqsecnum,aps,prb,twocolumn]{revtex}
\begin{document}
\baselineskip 0.5cm

\title{Magnetic precursor effects in Gd based intermetallic compounds }
\author{R. Mallik and E.V. Sampathkumaran }
\address{Tata Institute of Fundamental Research, Homi Bhabha Road, Mumbai - 400
005, INDIA. }
\maketitle

\begin{abstract}

The behaviour of electrical resistivity ($\rho$) and magnetoresistance in
the vicinity of respective magnetic ordering temperatures in a number of
Gd alloys is reported.  In some compounds, e.g., GdNi$_2$Sn$_2$ and
GdPt$_2$Ge$_2$, there is an enhancement of $\rho$ prior to long range
magnetic order over a wide temperature range which can be highlighted by
the suppression of $\rho$ caused by the application of a magnetic field.
However, such features are absent in many other Gd compounds, e.g.,
GdCu$_2$Ge$_2$, GdAg$_2$Si$_2$, GdAu$_2$Si$_2$, GdPd$_2$Ge$_2$ and
GdCo$_2$Si$_2$.  Attempts to relate such features to magnetic precursor
effects in heat capacity are made. On the basis of our studies, we suggest
that better understanding of  magnetic precursor effects in Gd alloys will
be helpful to throw light on some of the current trends in magnetism.
Various other interesting findings in the magnetically ordered state in
some of these alloys are also brought out. 

\end{abstract}
\vskip 1cm
PACS numbers:   
\par
72.15.C,E   (Electronic conduction in metals and alloys);
\par
75.20.H     (Local magnetic moment)
\par    
75.30.E     (Exchange interaction - magnetically ordered materials)

\section{Introduction}
One of the points of debate in the field of giant magnetoresistance (GMR)
is the origin of negative temperature coefficient of  resistivity ($\rho$)
above Curie temperature (T$_C$) and resultant large negative
magnetoresistance at T$_C$ [Ref. 1, 2].  Keeping such trends in the field
of magnetism in recent years in mind, we have been carefully investigating
the magnetoresistance behaviour of some of the Gd alloys in the vicinity
of respective magnetic ordering temperatures (T$_o$), in order to address
the question whether such features can arise from some other factor. We
have indeed noted an extra contribution to $\rho$ over a wide temperature
range above T$_o$ in GdPt$_2$Si$_2$, GdPd$_2$In,  GdNi$_2$Si$_2$ [Ref. 3],
GdNi [Ref. 4], and Gd$_2$PdSi$_3$ [Ref. 5], as a result of which the
magnetoresistance is negative just above T$_o$, attaining a large value at
T$_o$, similar to the behavior in manganites. In fact, in one of the Gd
compounds, Gd$_2$PdSi$_3$, the temperature coefficient of $\rho$ is even
negative just above N\'{e}el temperature (T$_N$), with a distinct minimum
at a temperature far above T$_N$. Such observations suggest the need to
explore the role of any other factor before long range magnetic order sets
in.  Similar resistance anomalies have been noted  above T$_o$ even in
some Tb and Dy alloys.\cite{6} Since critical spin fluctuations may set in
as one approaches T$_o$, the natural tendency is to attribute these
features to such spin fluctuations extending to unusually higher
temperature range.  In our opinion, there exists a more subtle effect,
e.g., a magnetism-induced electron localisation (magnetic polaronic
effect) and consequent reduction in the mobility of the carriers as one
approaches long range magnetic order.\cite{4,5,6}  The efforts on
manganites along these lines are actually underway and it appears that a
decrease in mobility of the carriers are primarily responsible for
negative temperature coefficient of $\rho$ above T$_C$ and large
magnetoresistance.\cite{7,8}
\par
The results on the Gd alloys  mentioned above are also  important to
various  developments in the field of heavy-fermions and Kondo lattices,
as discussed in Refs.  3-5, 9-11. Thus, the investigation of magnetic
precursor effects in relatively simple magnetic systems is relevant to
current trends in magnetism in general; the Gd systems are simple in the
sense that Gd does not exhibit any complications due to double-exchange,
crystal-fields, Jahn-Teller and Kondo effects.
\par
We therefore consider it worthwhile to get more experimental information
on magnetic precursor effects in Gd systems. In this article, we report
the results of electrical resistivity ($\rho$) measurements in a number of
other Gd alloys crystallizing in the same  (or closely related) structure,
in order to arrive at a overall picture of the magnetic precursor effects
in Gd compounds.  Among the Gd alloys investigated, interestingly, many do
not exhibit such resistance anomalies;  in addition, we find that there is
no one-to-one correspondence between the (non)observation of excess $\rho$
and a possible enhancement of heat capacity (C) above T$_o$ in these Gd
alloys.  The compounds\cite{12} under investigation are: GdCu$_2$Ge$_2$
(T$_N$= 12 K), GdAg$_2$Si$_2$ (T$_N$=  17 K), GdPd$_2$Ge$_2$ (T$_N$= 18 K,
Ref. 13), GdCo$_2$Si$_2$ (T$_N$= 44 K), GdAu$_2$Si$_2$ (T$_N$= 12 K),
GdNi$_2$Sn$_2$ (T$_N$= 7 K) and GdPt$_2$Ge$_2$ (T$_N$= 7 K). While the
crystallographic and  magnetic behaviour of most of  these compounds have
been well-known,\cite{12} this article reports  magnetic characterization
to our knowledge for the first time for GdNi$_2$Sn$_2$ and GdPt$_2$Ge$_2$.
We have chosen this set of compounds, since all of  these compounds are
crystallographically related: most of these form in ThCr$_2$Si$_2$-type
tetragonal structure, while GdNi$_2$Sn$_2$ and GdPt$_2$Ge$_2$ appear to
form in a related structure, viz., CaBe$_2$Ge$_2$ or its monoclinic
modification.\cite{14,15}

\section{Experimental}
\par
The samples were prepared by arc melting stoichiometric amounts of
constituent elements in an arc furnace  in an atmosphere of argon and
annealed at 800 C for 7 days. The samples were characterized by x-ray
diffraction. The electrical resistivity measurements were performed in
zero field as well as in the presence of a magnetic field (H) of 50 kOe in
the temperature interval 4.2 - 300 K by a conventional four-probe method
employing a silver paint for electrical contacts of the leads with the
samples; in addition, resistivity was measured as a function of H at
selected temperatures; no significance may be attached to the absolute
values of $\rho$ due to various uncertainties arising from the brittleness
of these samples, voids and the spread of silver paint. We also performed
the C measurements by a semiadiabatic heat-pulse method in the temperature
interval 2 - 70 K in order to look for certain correlations with the
behavior in $\rho$; respective non-magnetic Y or La compounds  have also
been measured so as to have an idea on the  lattice contribution, though
it is not found to be reliable at high temperatures (far above T$_o$). In
order to get further information on the magnetic behavior, the magnetic
susceptibility ($\chi$) was also measured in a magnetic field of 2 kOe (2
- 300 K) employing a superconducting quantum interference device and the
behavior of isothermal magnetization (M) was also obtained at selected
temperatures.

\section{Results and discussion}
\par
The results of $\rho$ measurements in the absence and in the presence of a
magnetic field are shown in Fig. 1a below 45 K for GdPt$_2$Ge$_2$. The C
data are shown in Fig. 1b. The $\chi$ data in the same temperature
interval are shown in Fig. 1c to establish the value of T$_N$.  The
magnetoresistance, defined as $\Delta\rho/\rho$=
[$\rho$(H)-$\rho$(0)]/$\rho$(0), as a function of H at selected
temperatures are shown in Fig. 1d. From the comparison of the data in
Figs. a, b and c, it is clear that this compound undergoes long range
magnetic ordering at (T$_N$= ) 7 K, presumably of an antiferromagnetic
type, considering that the Curie-Weiss temperature ($\theta$$_p$) obtained
from high temperature Curie-Weiss behavior of $\chi$ is negative (-8 K)
and the isothermal magnetization (M) at 4.5 K does not show indication for
saturation (and, in fact, varies linearly with H, Fig.  1c, inset).  There
is an upturn in $\rho$ below 7 K, instead of a drop, presumably due to the
development of magnetic Brillouin-zone boundary gaps;\cite{16} However,
with the application of a magnetic field, say 50 kOe,  this low
temperature upturn in $\rho$  gets depressed; the point to be noted is
that there is a significant depression of $\rho$ with the application of H
even above 7 K, the magnitude of which decreases with increasing
temperature.  Thus, there is a significant negative magnetoresistance not
only below T$_N$, but also above it over a wide temperature range. This
point can be emphasized more clearly when one measures  $\Delta\rho/\rho$
as a function of H at various temperatures (Fig.  1d). There is  a
quadratic variation with H (up to about 50 kOe) at all temperature
mentioned in the plots, attaining a large value at higher fields, and
these are characteristics of spin-fluctuation systems.  In order to
explore whether  any such magnetic precursor effects are present in the C
data, we show the magnetic contribution (C$_m$) to C in Fig. 1b after
subtracting the lattice contribution (derived from the C data of
YPt$_2$Ge$_2$) as described in Refs.  9, 17.  It appears that this may not
be the perfect way of determination of C$_m$ above 30 K as the derived
lattice part does not coincide with the measured data for the sample,
though the magnetic entropy (obtained by extrapolation of C$_m$ to zero
Kelvin) reached its highest value (R ln 8) around 40 K; there may possibly
be different degree of crystallographic disorder between Gd and Y alloys,
which is responsible for this discrepancy.  Clearly the feature is rounded
off at the higher temperature side of T$_N$, resulting in a tail extending
to higher temperature range and this feature is free from the error
discussed above.  The data basically provide evidence for the fact that
the full magnetic entropy (R ln8) is attained only in the range 30 - 40 K
and it is exactly the same temperature range till which we see an
enhancement of $\rho$, depressing with the application of H.  In short,
this compound exhibits magnetic precursor effects both in C and $\rho$
data.
\par     
As in the case of GdPt$_2$Ge$_2$, the results obtained from various
measurements for GdNi$_2$Sn$_2$ are shown in Fig. 2 below 35 K. It is
clear from the features in $\rho$, C and $\chi$ that this compound orders
magnetically at about 7 K; from the reduced value of peak C$_m$ (lattice
contribution derived from the values of YNi$_2$Sn$_2$) [Ref. 17]  and
negative $\theta$$_p$, we infer that the magnetic structure is of an
amplitude-modulated antiferromagnetic type.  The main point of emphasis is
that there is an excess resistivity till about 15 K, which is highlighted
by the depression of $\rho$ with the application of H. Though there are
problems similar to GdPt$_2$Ge$_2$ in deducing precise lattice
contribution at higher temperature,  we are confident that C$_m$ data
(qualitaively) exhibit a tail till about 15 K and the total magnetic
entropy is released around the same temperature. The magnetoresistance
appears to vary nearly quadratically with H above T$_N$, say, at 10 and 15
K.  Thus, $\rho$ and C data show magnetic precursor effects for this alloy
as well.
\par
We now present the results on a series of Gd alloys in which the excess
resistance (in the sense described above) is not observable above T$_o$.
These alloys are GdCo$_2$Si$_2$ (Fig. 3), GdAu$_2$Si$_2$ (Fig. 4) and
GdPd$_2$Ge$_2$ (Fig. 5). It is clear from the figures 3-5 that the
resistivity  in the presence and in the absence of H are practically the
same (within 0.1\%) above their respective ordering temperatures, thereby
establishing the absence of an additional contribution to $\rho$ before
long range ordering sets in. In order to look for the 'tail' in C$_m$
above T$_o$, we attempted to obtain respective lattice contributions
(employing the C values of YCo$_2$Si$_2$, YAu$_2$Si$_2$ and YPd$_2$Ge$_2$
respectively). We can safely state that the continuous decrease in C$_m$
just above T$_o$, if exists, does not proceed beyond 1.2T$_o$  (See Figs.
3b, 4b and 5b).  Thus, it appears that the magnetic precursor effects in
C,  if present, are negligible, thus tracking the behavior of "excess
resistance".
\par
In GdCu$_2$Ge$_2$ and GdAg$_2$Si$_2$ as well, clearly there is no excess
resistivity above T$_N$, as the application of H does not suppress the
value of $\rho$ (Figs. 6 and 7). However, it contrast to the cases
discussed in the previous paragraph, it appears that there is no
correlation between C and $\rho$ behavior prior to long range magnetic
order. YCu$_2$Ge$_2$ and LaAg$_2$Si$_2$ have been used as references to
obtain lattice contributions to C respectively. The finding of interest is
that the magnetic contribution to C appears to exhibit a prominent tail
(without any doubt in GdCu$_2$Ge$_2$), at least till 10 K above respective
T$_N$. This behavior is similar to that noted for GdCu$_2$Si$_2$
earlier.\cite{11,17}
\vskip 0.5cm
{\it We have also made various other interesting findings:} 
\vskip 0.5cm
\par
The peak values of C$_m$ for GdPt$_2$Ge$_2$ and GdNi$_2$Sn$_2$ are much
smaller than that expected (20.15 J/mol K, Ref. 17) for equal moment
(simple antiferro, ferro or helimagnetic) magnetic structures and the fact
that the value is reduced by at least a factor of about 1/3 shows that the
magnetic structure is modulated.  The situation is somewhat similar for
GdPd$_2$Ge$_2$. However, for GdCo$_2$Si$_2$ and GdAu$_2$Si$_2$, the peak
values of C$_m$ are very close to the expected value for commensurate
magnetic structures, thus suggesting that the (antiferromagnetic) magnetic
structure is not modulated.
\par
For GdNi$_2$Sn$_2$ (Fig. 2d), $\Delta\rho/\rho$  as a function of H at 4.5 K
exhibits a sharp rise for initial applications of H with a positive peak
near 8 kOe. While the positive sign may be consistent with
antiferromagnetism, corresponding anomaly in the isothermal magnetization
at 4.5 K is not very prominent; the plot of M versus H, however, is not
perfectly linear at 4.5 K, showing a weak metamagnetic tendency around
30 kOe (Fig. 2c, inset). It appears that the peak in the magnetoresistance
is a result of significant changes in the scattering effects from a weak
metamagnetism. Even in the case of GdAg$_2$Si$_2$, there is a weak feature
in the plot of M vs H at 4.5 K around 40 kOe due to possible metamagnetic
transition (see Fig. 7d), which is pronounced in the magnetoresistance
beyond 20 kOe. In the case of GdPd$_2$Ge$_2$, at 5 K, $\Delta\rho/\rho$
shows a positive value till 20 kOe, beyond which the value is negative
exhibiting a non-monotonic variation with H (Fig. 5d); the plot of M
versus H shows only a small deviation from linearity around this field.
Thus there are very weak metamagnetic effects which have subtle effects on
the scattering processes in the antiferromagnetically ordered state in
these compounds.  The plot of magnetoresistance versus H and that of
isothermal magnetization look similar for GdCu$_2$Ge$_2$ (Figs. 6d), with
a very weak metamagnetic tendency near 35 kOe, as reflected by non-linear
plots.  These results suggest that the magnetoresistance technique is a
powerful tool to probe metamagnetism, even the weak ones, which may not be
clearly detectable by magnetization measurements.
\par
It is to be noted that, interestingly, the value of magnetoresistance is
very large at high fields at 5 K (see Fig. 7d) for GdAg$_2$Si$_2$
(possibly due to granularity?). The heat capacity data in the magnetically
ordered state in GdAg$_2$Si$_2$ as well as in GdPd$_2$Ge$_2$, reveal the
existence of additional shoulders, which may be the result of a combined
influence of spin reorientation  and Zeeman effects.\cite{17} In
particular, for GdAg$_2$Si$_2$, the magnetic behavior appears to be
complex due to the presence of two prominent magnetic transitions,
(interestingly) a discontinuous one near  17 K and the other at 11 K (see
the features in C and $\chi$ in Figs. 7b and 7c).  At the 17K-transition
in this compound, there is a sudden upward jump in C, and at the same
temperature $\rho$ shows a sudden upturn instead of a decrease (Fig. 7),
possibly due to the formation of antiferromagnetic energy gaps.  It would
be interesting to probe whether the transition is first-order in nature.
It appears that there is another magnetic transition in GdCo$_2$Si$_2$ as
well around 20 K as seen by an upturn in the susceptibility (Fig. 3c).
\section{CONCLUSIONS}
\par 
To summarise, on the basis of  our investigations on Gd alloys, we divide
the Gd compounds into two classes: {\bf Class I}, in which there is an
excess contribution to $\rho$ prior to long range magnetic order over a
wide temperature range, as a result of which the magnetoresistance is
large and negative, e.g., GdNi, GdNi$_2$Si$_2$, GdPt$_2$Si$_2$,
GdPt$_2$Ge$_2$.  GdNi$_2$Sn$_2$, GdPd$_2$In, Gd$_2$PdSi$_3$. {\bf Class
II}, in which such features are absent, e.g., GdCu$_2$Si$_2$,
GdCu$_2$Ge$_2$, GdAg$_2$Si$_2$, GdAu$_2$Si$_2$, GdCo$_2$Si$_2$,
GdPd$_2$Ge$_2$. (At this juncture, we would like to add that we performed
similar studies on compounds like, GdCu$_2$, GdAg$_2$, GdAu$_2$,
GdCoSi$_3$   and GdNiGa$_3$ and we do not find any magnetic precursor
effects).  The present study on isostructural compounds establishes that
there is no straightforward relationship between the observation of the
excess $\rho$ on the one hand and the crystal structure or the type of
transition metal and s-p ions present in the compound on the other. The
fact that all the compounds studied in this investigation are of layered
type suggests that possible onset of magnetic correlations within a layer
before long range magnetic order sets in cannot be offerred as the sole
reason for excess resistivity selectively in some cases. If one is tempted
to attribute the observation of excess $\rho$ to critical spin
fluctuations extending to higher temperature range, as inferred from the
tail in C$_m$ above T$_o$, one does not get a consistent picture, the
reason being that, in some of the class II alloys, there is a distinct
tail in C$_m$. It is therefore clear that there must be more physical
meaning for the appearance of excess $\rho$ in class I alloys. As proposed
in Refs 5 and 6, one may have to invoke the idea of "magnetic disorder
induced localisation of electrons" (and consequent reduction in the
mobility of the charge carriers) before the onset of long range order, as
a consequence of short range magnetic order, detected in the form of a
tail in heat capacity. The data presented in this article essentially
demand that one should explore various factors determining the presence or
the absence of the proposed "magnetic-localisation" effects in the
presence of short-range magnetic correlations; possibly, the relative
magnitudes of mean free path, localisation length\cite{7} and short range
correlation length play a crucial role. It is worthwhile to pursue this
question, so as to throw light on several issues in current trends in
magnetism.

 \begin{figure}  

 \caption{(a) Electrical resistivity in zero field and in the presence of
a magnetic field of 50 kOe, (b) Heat capacity (C), lattice contribution to
C and the magnetic contribution (C$_m$) to C and (c) the magnetic
susceptibility below 45 K as well as the isothermal magnetization (inset)
at 5 K for GdPt$_2$Ge$_2$. The magnetoresistance, $\Delta\rho/\rho$, as
a function of magnetic field (H) at various temperatures is shown in Fig.
(d). The lines drawn through the data points serve as guides to the eyes.}

 \end{figure}

\begin{figure}  

 \caption{(a) Electrical resistivity in zero field and in the presence of
a magnetic field of 50 kOe, (b) Heat capacity (C), lattice contribution to
C and the derived magnetic contribution (C$_m$) to C and (c) the magnetic
susceptibility for GdNi$_2$Sn$_2$ below 32 K. The magnetoresistance,
$\Delta\rho/\rho$, as a function of magnetic field (H) at various
temperatures is shown in Fig.  (d). The lines drawn through the data
points serve as guides to the eyes.  The isothermal magnetization behavior
at 4.5 K is plotted in the inset of figure (c) and the low field linear
region is shown by a continuous line.}

\end{figure}

\begin{figure}  

 \caption{(a) Electrical resistivity in zero field and in the presence of
a magnetic field of 50 kOe, (b) Heat capacity (C), lattice contribution to
C and the magnetic contribution (C$_m$) to C and (c) the magnetic
susceptibility below 60 K as well as the isothermal magnetization (inset)
at 4.5 K for GdCo$_2$Si$_2$. The lines drawn through the data points serve
as guides to the eyes.}

 \end{figure}

\begin{figure}  

\caption{(a) Electrical resistivity in zero field and in the presence of
a magnetic field of 50 kOe, and (b) Heat capacity (C), lattice
contribution to C and the magnetic contribution (C$_m$) to C for
GdAu$_2$Si$_2$ below 30 K.} 

\end{figure}

\begin{figure}  

 \caption{(a) Electrical resistivity in zero field and in the presence of
a magnetic field of 50 kOe, (b) Heat capacity (C), lattice contribution to
C and the magnetic contribution (C$_m$) to C and (c) the magnetic
susceptibility below 35 K for GdPd$_2$Ge$_2$.  The magnetoresistance,
$\Delta\rho/\rho$, and isothermal magnetizaton as a function of magnetic
field (H) at 5 K are shown in Fig. (d). The lines drawn
through the data points serve as guides to the eyes in all the plots
except for the M versus H plot, in which case the straight line represents
the low field linear region.}

 \end{figure}

\begin{figure}  

 \caption{(a) Electrical resistivity in zero field and in the presence of
a magnetic field of 50 kOe, (b) Heat capacity (C), lattice contribution to
C and the magnetic contribution (C$_m$) to C and (c) the magnetic
susceptibility below 30 K for GdCu$_2$Ge$_2$.  The magnetoresistance,
$\Delta\rho/\rho$, and isothermal magnetization as a function of magnetic
field (H) at 4.5 K are plotted in Fig. (d). The lines drawn
through the data points serve as guides to the eyes.}

 \end{figure}

\begin{figure}  

\caption{(a) Electrical resistivity in zero field and in the presence of a
magnetic field of 50 kOe, (b) Heat capacity (C), lattice contribution to C
and the magnetic contribution (C$_m$) to c and (c) the magnetic
susceptibility below 30 K for GdAg$_2$Si$_2$.  The magnetoresistance,
$\Delta\rho/\rho$, and isothermal magnetization as a function of magnetic
field (H) at 4.5 K are plotted in Fig. (d). The lines drawn through the
data points serve as guides to the eyes.}

 \end{figure}

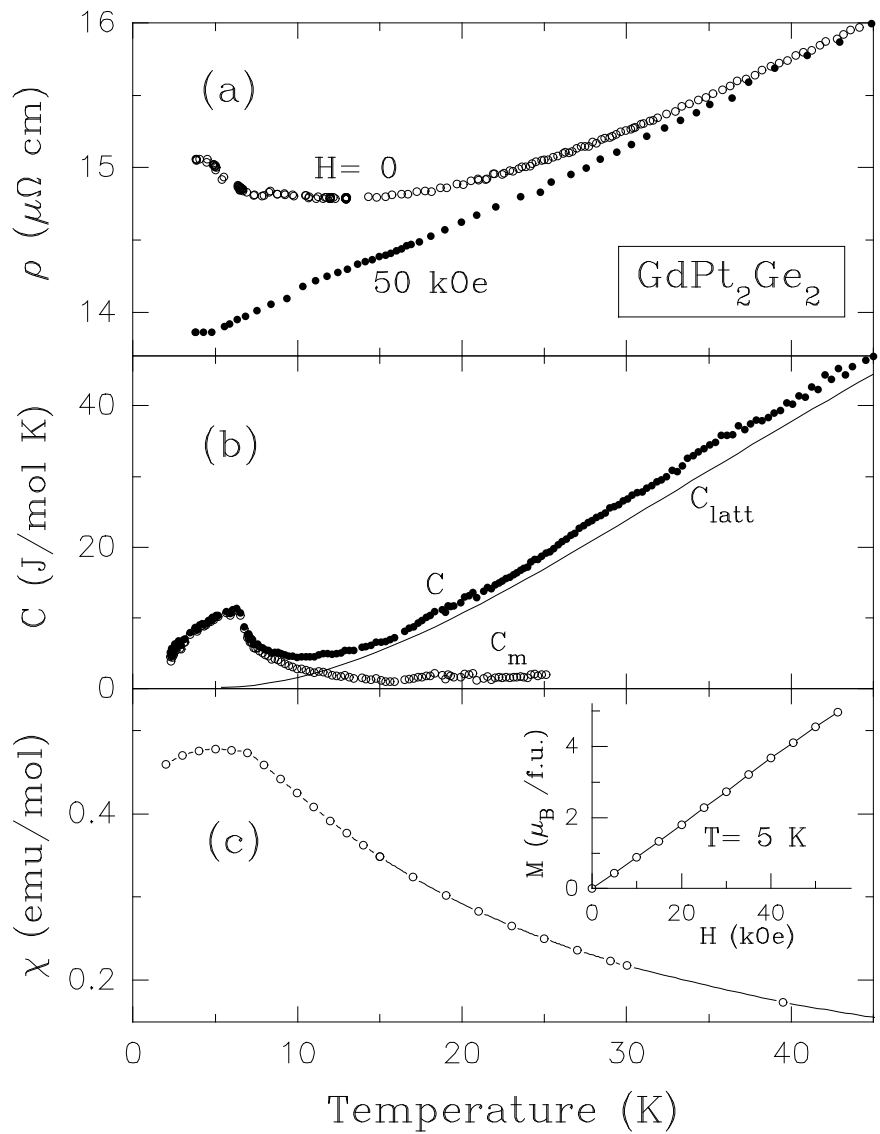
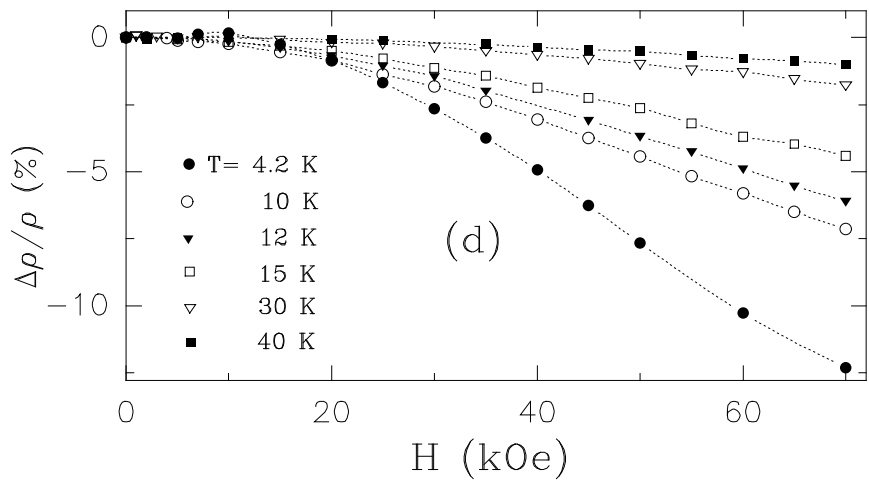
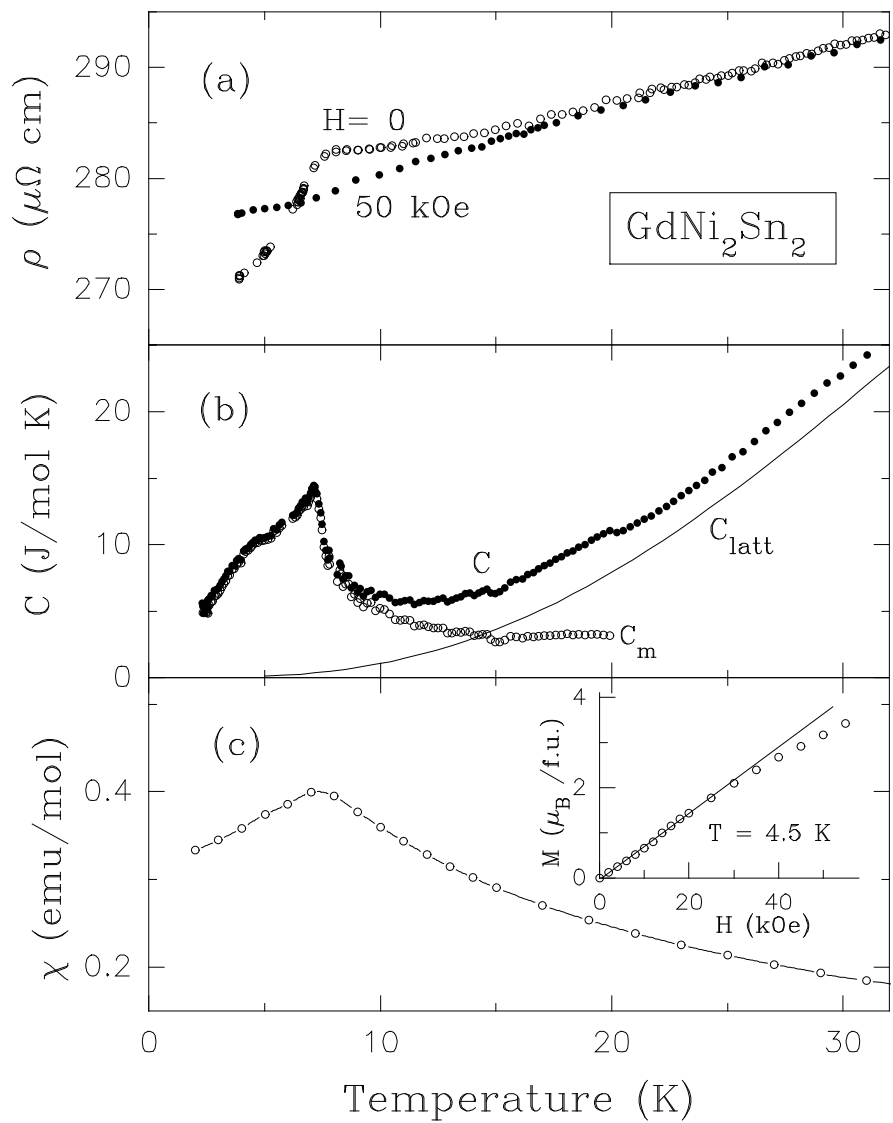
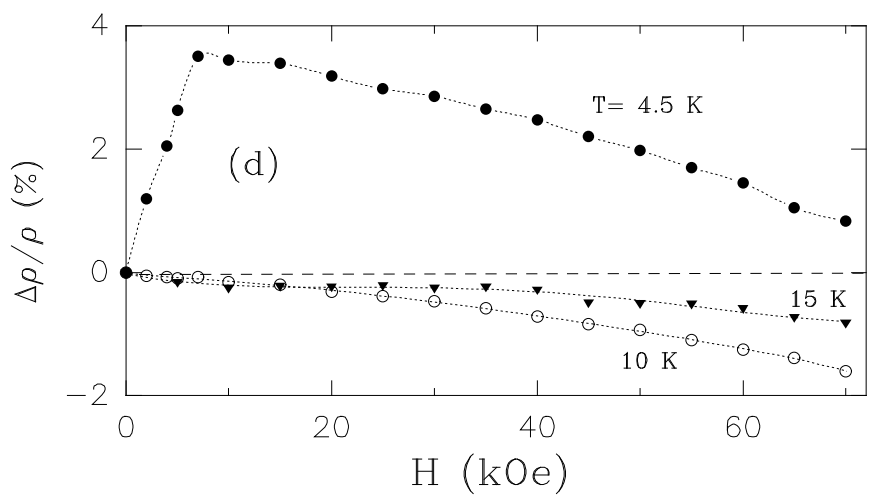
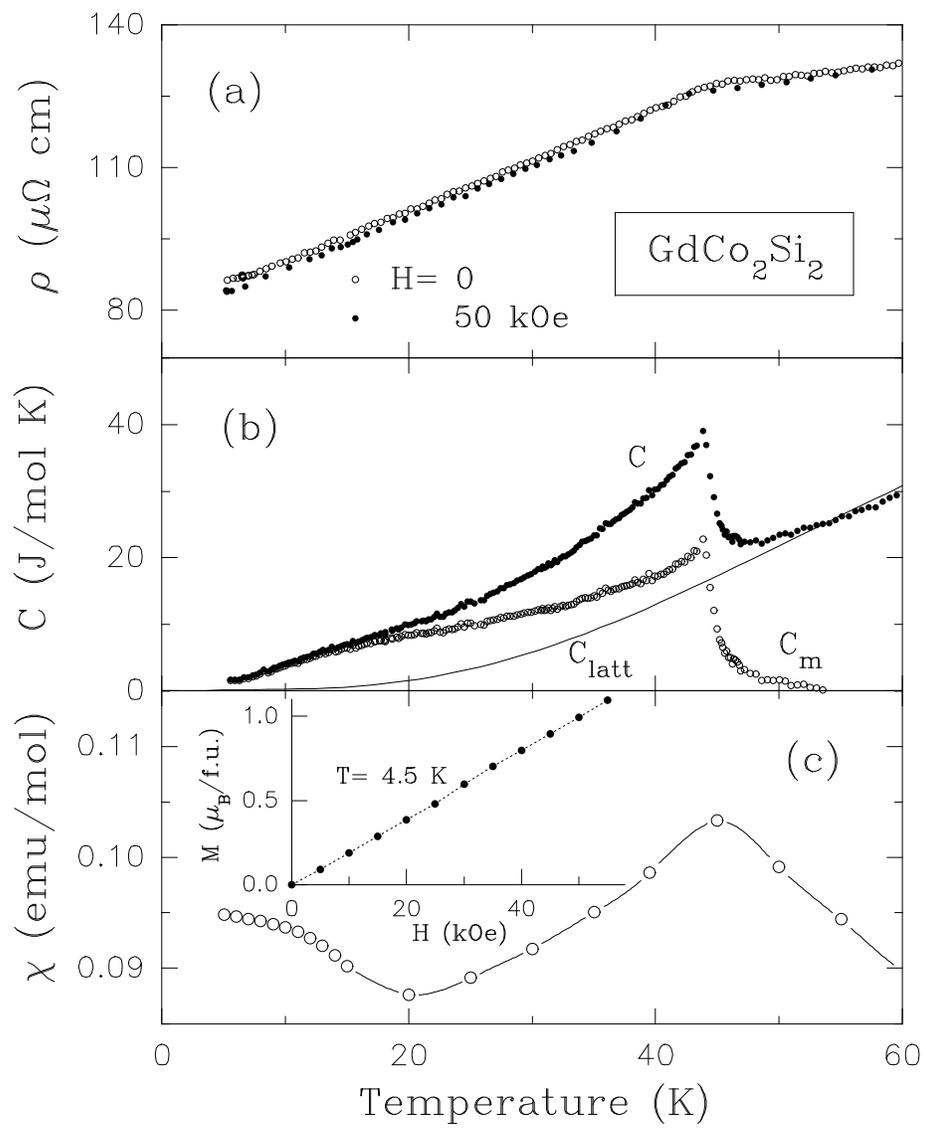
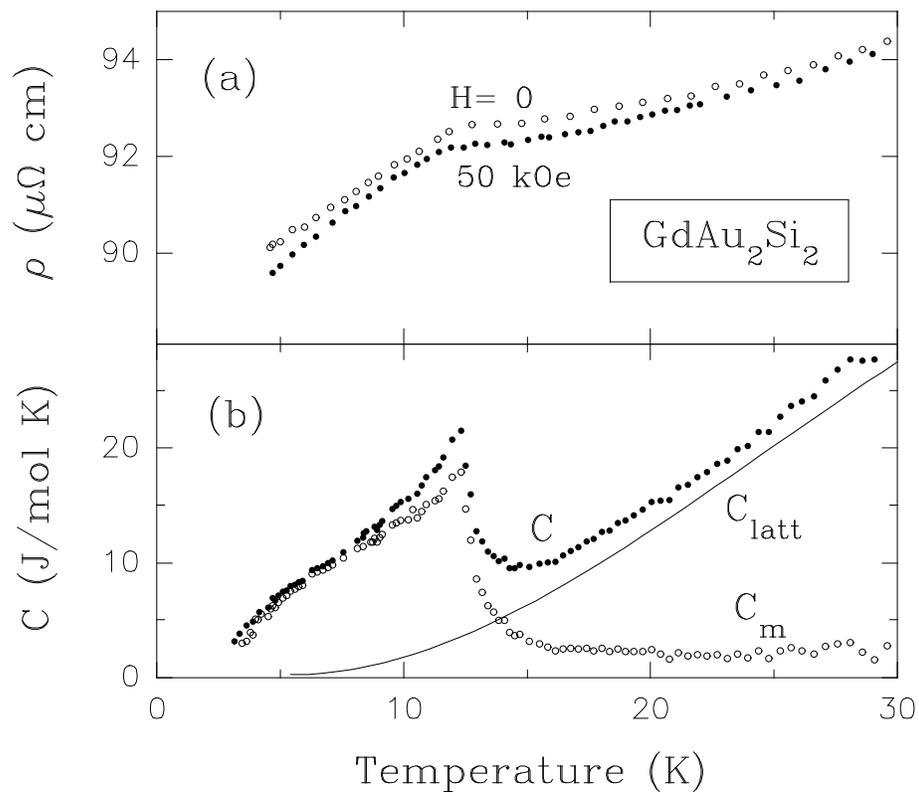
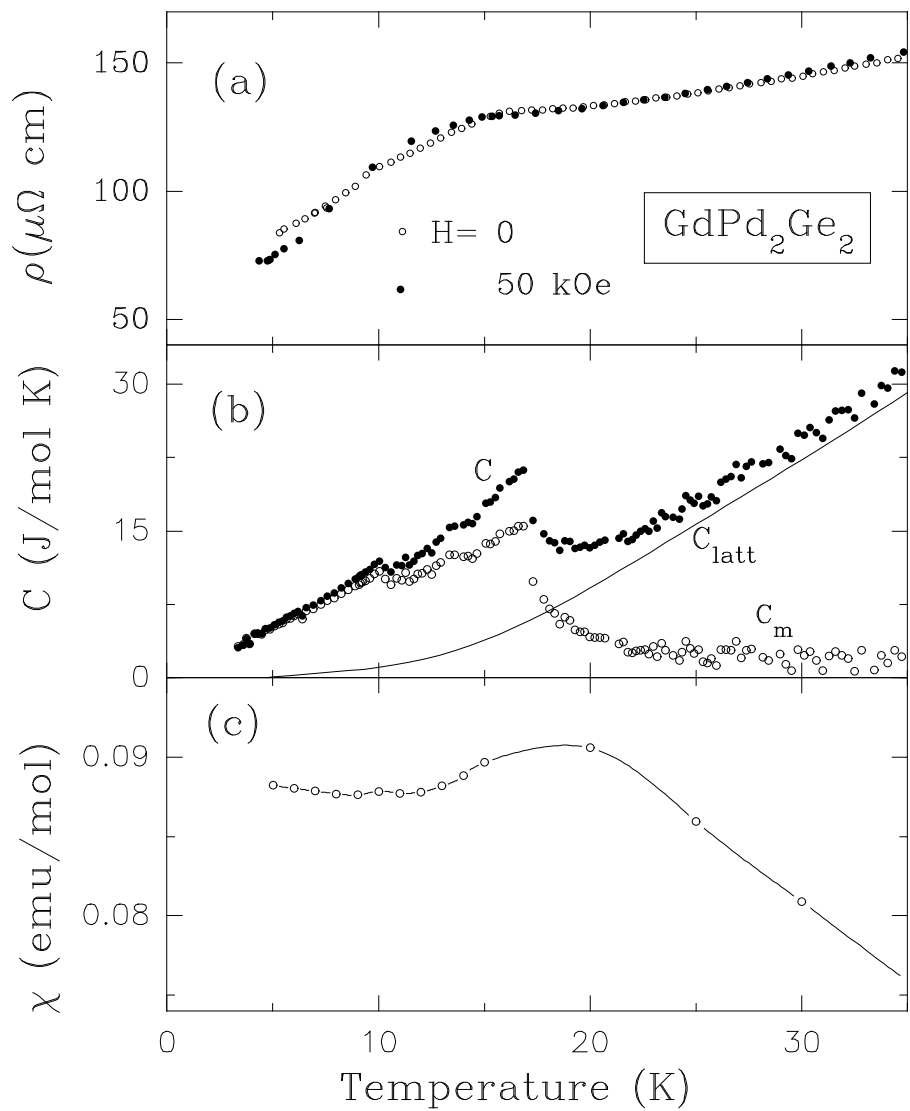
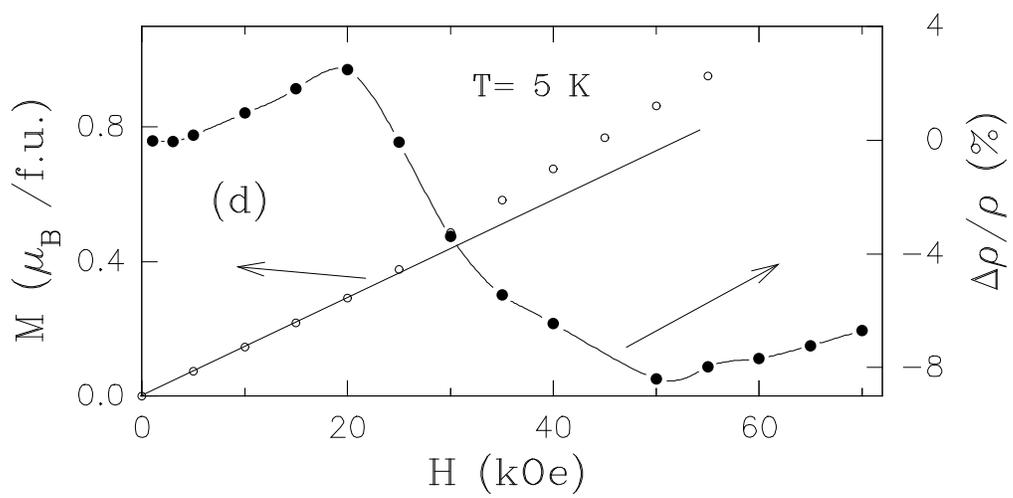
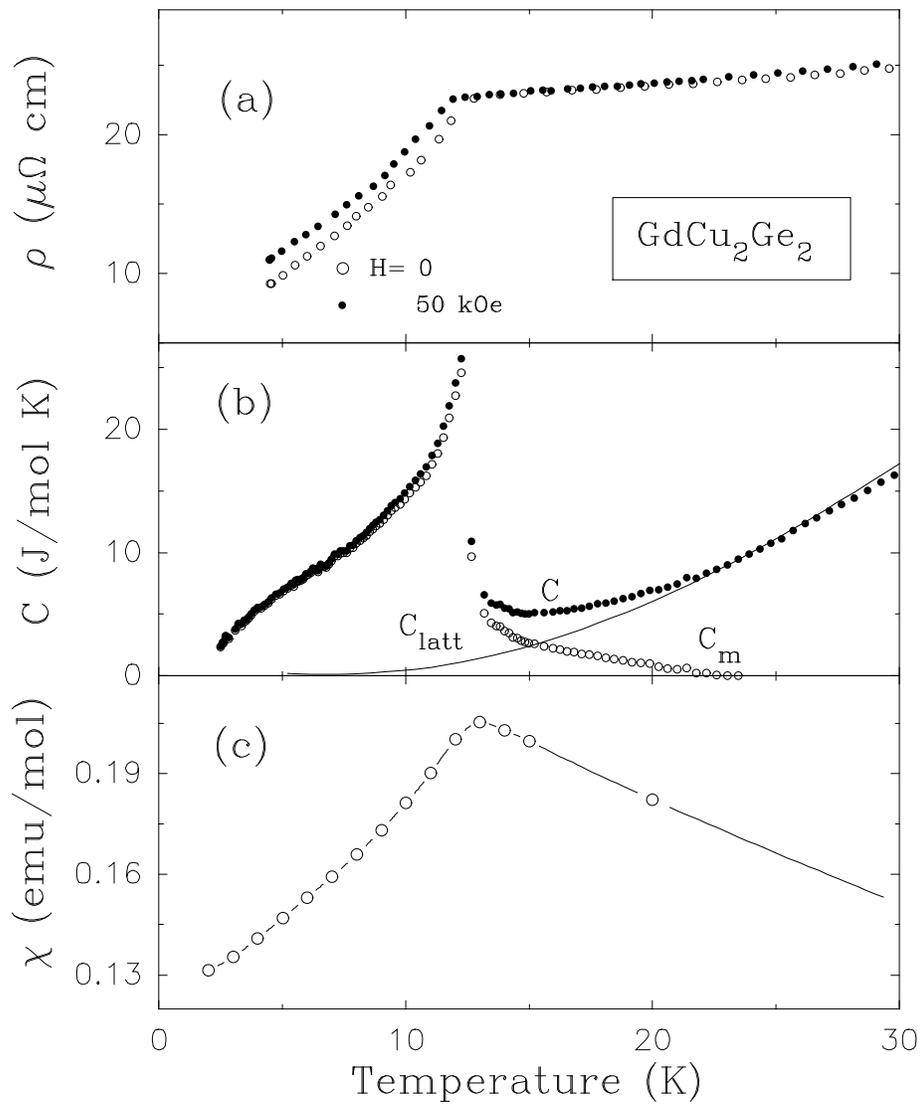
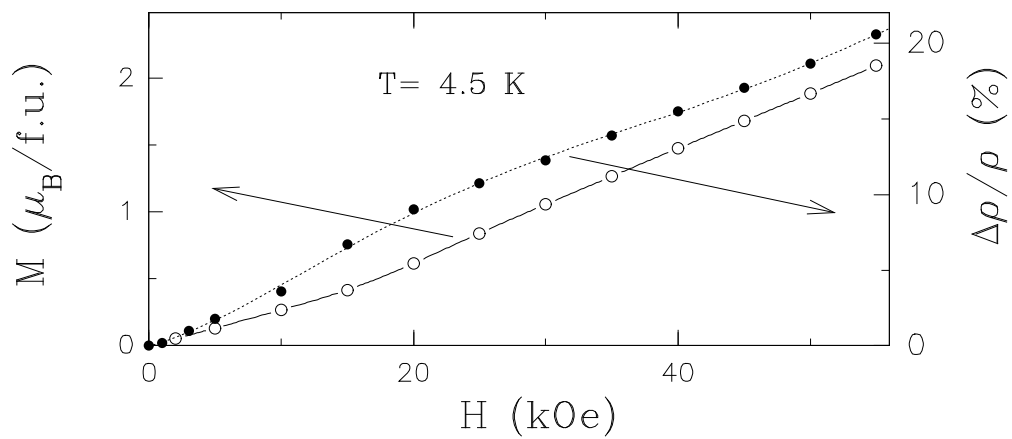
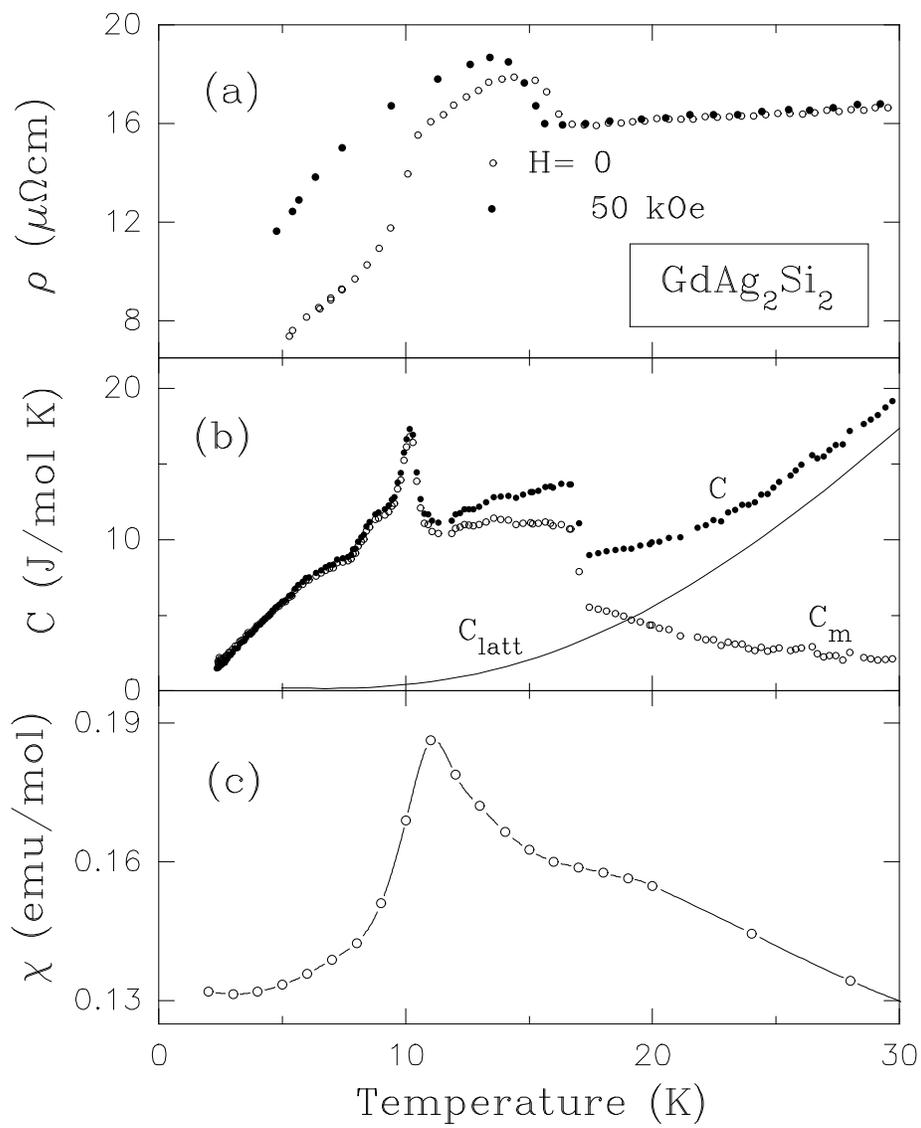
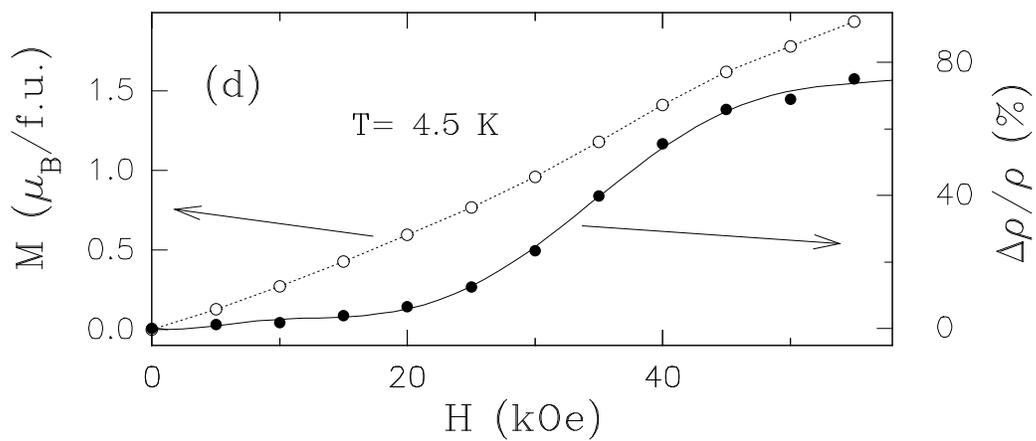
\end{document}